\documentclass[12pt]{iopart}

\expandafter\let\csname equation*\endcsname\relax 
\expandafter\let\csname endequation*\endcsname\relax 

\usepackage{amssymb}
\usepackage{mathtools}
\usepackage{harpoon}

\usepackage{xcolor}
\usepackage{graphicx} % Include figure files
\begin{document}

\title[A New Beam Neutron Lifetime Experiment in Superfluid Helium-4]{A New Neutron Lifetime Experiment with Cold Neutron Beam Decay in Superfluid Helium-4}

\author{Wanchun Wei}
\ead{wanchun.wei@caltech.edu} 
\vspace{10pt}
\address{Kellogg Radiation Laboratory, California Institute of Technology, Pasadena, California 91125, USA}
\begin{indented}
\item[]May 2020
\end{indented}

\begin{abstract}
The puzzle remains in the large discrepancy between neutron lifetime measured by the two distinct experimental approaches -- counts of beta decays in a neutron beam and storage of ultracold neutrons in a potential trap, namely, the beam method versus the bottle method. In this paper, we propose a new experiment to measure the neutron lifetime in a cold neutron beam with a sensitivity goal of 0.1\% or sub-1 second. The neutron beta decays will be counted in a superfluid helium-4 scintillation detector at 0.5~K, and the neutron flux will be simultaneously monitored by the helium-3 captures in the same volume. The cold neutron beam must be of wavelength $\lambda>$16.5~\AA~to eliminate scattering with superfluid helium. A new precise measurement of neutron lifetime with the beam method of unique inherent systematic effects will greatly advance in resolving the puzzle.
\end{abstract}

\vspace{2pc}

\submitto{\JPG}

%\maketitle

\section{Introduction}
A precise measurement on the neutron lifetime is {\color{black}important} to many fundamental questions in particle physics, astrophysics and cosmology, such as CKM unitarity and primordial helium abundance in Big Bang Nucleosynthesis (BBN).\cite{Wietfeldt11, Belfatto20, Dubbers11} So far, its values obtained from the two distinct methods significantly differ from each other{\color{black}\cite{nLifetimeDiff1, nLifetimeDiff2, Wietfeldt18}, possibly due to unaccounted systematic effects in either or both of the methods; yet otherwise it implies new physics\cite{Mirror19_1, Mirror19_2, NP_1, NP_2, NP_3, NP_4, NP_5, NP_6}, many theories of which remain controversial\cite{Czarnecki18, Dubbers19,Berezhiani19}.} On one side, the measurement is done in a neutron beam by counting the number of neutrons undergoing beta decay when the neutron flux passes through a defined volume. It is thus called the beam method. The weighted average of the recent two beam lifetime measurements with a proton quasi-Penning trap in a cold neutron (CN) beam is $\tau_n = 888.0 \pm 2.0$~s.\cite{Byrne96, Yue13} On the other side, ultracold neutrons (UCN) can be stored in a material box or magneto-gravity trap, and the neutron lifetime is measured by counting the surviving neutrons after a period of storage. It is called the bottle method. The weighted average of several recent bottle lifetime measurements is $\tau_n = 879.4 \pm 0.6$~s.\cite{Serebrov05, Pichlmaier10, Steyerl12, Arzumanov15, Serebrov18, Pattie18, Ezhov18} The difference is as large as $8.7 \pm 2.1$~s ($4.1\sigma$). Many further experimental efforts are on the way to address the discrepancy. While existing experiments are upgrading to improve their statistics and searching for hidden systematic effects, new experimental strategies with a distinct set of systematic effects are being proposed and carried out. For instance, researchers in J-PARC started a new measurement in the pulsed CN beam by characterization of the electron recoils in the beta decay events and the helium-3 capture events in a Time Projection Chamber (TPC) filled with gaseous mixture of helium and carbon dioxide.\cite{Nagakura17,Hirota20} It is a revival of the beam experiment originally proposed by Kossakowski {\it et al.} in 1989 \cite{Kossakowski89}. Researchers at Los Alamos National Laboratory are prototyping a beam/bottle hybrid experiment, named UCNProBe, to measure the number of decays and helium-3 captures via detection of scintillation in a UCN storage box.\cite{Tang19} 

In this paper, we propose a new experimental method with a different combination of existing technologies, in order to resolve the neutron lifetime enigma\cite{Greene16}. The proposed experiment is essentially a beam lifetime measurement. It counts the decay product -- electrons, rather than protons, via detection of electron recoil scintillation in superfluid helium-4 at 0.5~K. In order to eliminate neutron scattering with superfluid helium, the CN beam must be of wavelength $\lambda >16.5~\mbox{\AA}$, where kinematics of scattering can never be satisfied.\cite{Cohen57} The neutron flux is monitored by helium-3 captures via nuclear recoil scintillation in the same volume of superfluid helium-4. The decay events may be distinguished from the capture events, as the features of scintillation differ between the electron and nuclear recoils. Meanwhile, a precise beta spectrum of neutrons, in addition to the neutron capture peak, will be constructed in a wide energy window with good resolution. A fit of beta spectrum can separate the overlapping counts of decay events from capture events, and complement the total counts with the missing number of electrons outside the detectable energy window. In the end, an accurate neutron lifetime can be obtained with a good knowledge of the ratio of capture-to-decay event rates and the helium-3 density in superfluid helium-4.

\section{Experimental Method}
\label{sec::expMethod}

Suppose the decay volume is cylindrical with a length $L$ of 75~cm and a diameter $D$ of 7.5~cm, {\it i.e.} the length to diameter ratio is $L/D=10$. The CN beam of 3~cm diameter passes the decay volume along the axis of the cylinder. The detectable neutron decay rate in the volume is given as 
\begin{equation}
    \dot{N}_{\beta}=\tau_{\beta}^{-1}\epsilon_{\beta} L \int_{A_{b}} da \int_{v} dv~ I(v,\overrightharp{r})\frac{1}{v}
\end{equation}

where $\tau_{\beta}$ is the neutron lifetime, $\epsilon_{\beta}$ is the detection efficiency of the beta decay in the given geometry, $A_b$ is the cross sectional area of the beam, and $I(v,\overrightharp{r})$ is cold neutron fluence rate with respect to the neutron velocity $v$ and cross sectional distribution 
{\color{black} of positions} $\overrightharp{r}$.

When the CN beam passes through the decay volume, the $^3$He nuclei in superfluid helium-4 capture neutrons via nuclear reaction $n + {^3He} \rightarrow p + t + 764~\mbox{keV}$. The detectable capture rate is given as 

\begin{equation} 
    \dot{N}_{p+t} =\epsilon_{\scriptscriptstyle He3} \sigma^{th}_{\scriptscriptstyle He3} v^{th}_n n_{\scriptscriptstyle He3} L \int_{A_{b}} da \int_{v} dv~I(v,\overrightharp{r}) 
    \frac{1}{v}, 
\end{equation}

where $\epsilon_{\scriptscriptstyle He3}$ is the detection efficiency of the capture events, $\sigma^{th}_{\scriptscriptstyle He3}$ is the absorption cross section of $^3$He nuclei for thermal neutrons at a velocity $v^{th}_n=2200$~m~s$^{-1}$, and $n_{\scriptscriptstyle He3}$ is the $^3$He density. The neutron lifetime $\tau_{\beta}$ can be obtained from the ratio of the observed neutron $^3$He capture rate to the beta decay rate. 

\begin{equation} \label{eqn:nTauEqu}
     \tau_{\beta} = \frac{\dot{N}_{p+t}}{\dot{N}_{\beta}} \cdot \frac{\epsilon_{\beta}}{\epsilon_{\scriptscriptstyle He3}} \cdot  \frac{1}{\sigma^{th}_{\scriptscriptstyle He3} v^{th}_n n_{\scriptscriptstyle He3} }
\end{equation} 

Eqn. (\ref{eqn:nTauEqu}) is the key expression in this experiment. It explicitly shows the measurement of $\tau_{\beta}$ is independent of the neutron flux as well as the geometry of the decay volume. The overall accuracy relies on that of the observed ratio of event rates $\kappa =\dot{N}_{p+t}/\dot{N}_{\beta}$, the helium-3 density $n_{\scriptscriptstyle He3}$ in superfluid helium-4, and the detection efficiency of scintillation events $\epsilon_{\scriptscriptstyle He3}$ and $\epsilon_{\beta}$. The former two quantities will be experimentally acquired, and the detection efficiencies will be determined through simulations considering the calibration and background discrimination. 

\begin{figure}
  \centering
  \includegraphics[width=5.5in]{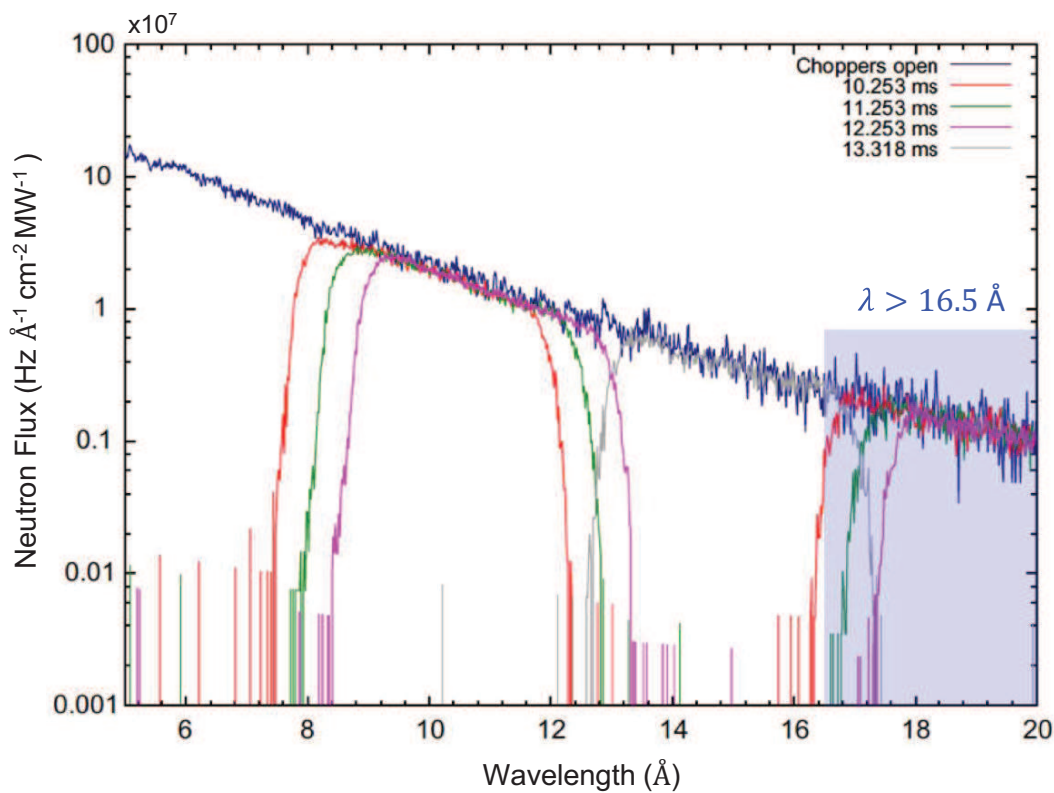}
  \caption{\label{fig:fig_FNPB} Spectrum of cold neutron beam at the SNS FnPB beamline with choppers (a reprint of Figure 5 in Fomin 2015 \cite{Fomin15}). The portion of wavelength $\lambda>16.5$~\AA~is highlighted in blue shadows. }
\end{figure}

Here, we provide an estimate of count rates based on the published performance of the Fundamental Neutron Physics Beam Line (FnPB) in the Spallation Neutron Source (SNS) at Oak Ridge National Laboratory (ORNL), as shown in Fig. \ref{fig:fig_FNPB} \cite{Fomin15}. The neutron flux at 17~\AA~is about $2.4\times10^6~$Hz~\AA$^{-1}$cm$^{-2}$MW$^{-1}$. With a time-averaged proton power of 1.8MW at 60Hz of double-chopper, the incident rate of 17~\AA~neutrons with 0.5~\AA~pulse width is about $1.5\times10^7$~Hz. \iffalse {\color{blue} (\missingcommand{CN_incident_rate})} \fi In this estimate, neutron lifetime is taken as the PDG suggested value $\tau_{\beta}=880$~s \cite{PDG18}. It takes about 3.2~ms for the 17~\AA~neutrons to pass 75~cm long decay volume, and the neutron decay probability is $3.66\times10^{-6}$ \iffalse {\color{blue} (\missingcommand{CN_beta_rate})} \fi for a CN beam with a cross section of 3~cm diameter. There are an average of 55.9 Hz \iffalse {\color{blue} (\missingcommand{CN_dN_beta_dt})} \fi of neutron decay events. The natural abundance of $^3$He in liquid helium is $X_{\scriptscriptstyle He3}=5\times10^{-7}$ in fractional concentration. Near isotopically pure $^4$He with $X_{\scriptscriptstyle He3}<2.5\times10^{-13}$ has been produced as reported by Hendry and McClintock.\cite{McClintock87} Assuming superfluid helium-4 with $X_{\scriptscriptstyle He3}=2\times10^{-10}$ \iffalse {\color{blue} (\missingcommand{X3})} \fi can be prepared, an average of 252.1 Hz \iffalse {\color{blue} (\missingcommand{CN_dN_capt_dt})} \fi neutron capture events will occur simultaneously when the CN beam passes the decay volume. 

\begin{figure}
  \centering
  \includegraphics[width=6in]{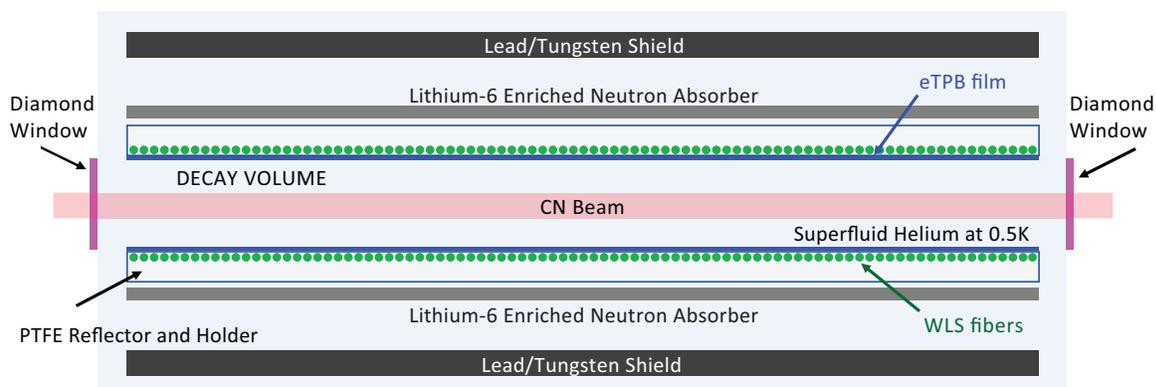}
  \caption{\label{fig:fig_schematic_detector} Schematic of the conceptual detector (non-scaled)}
\end{figure} 

% Section of scintillation signals in liquid helium
\section{Scintillation Signals in Liquid Helium}

The number of beta decays and neutron captures will be counted via scintillation signals in liquid helium. Liquid helium is an ideal scintillator that has been proposed and studied for detection of neutrino \cite{Adams00} and dark matter \cite{Guo13,Hertel19}. Compared to those experiments, the detection volume in the current beam lifetime experiment is compact. High detection efficiency of scintillation photons can be practically achieved, as well as a wide energy window with good resolution and small lower bound. 

In a neutron beta decay event, scintillation is generated by the recoiling electron, as the counterpart proton is too slow. About 35\% of the total energy in each electron recoil above 1~keV creates He$_2^*$ molecules of excited singlet state He$_2$($A^1\Sigma_u^+$) in liquid helium. The singlets radiatively decay in less than 10~ns and emit about 22 extreme ultraviolet (EUV) photons per keV of electron recoil energy {\it KE}$_e$, with a spectrum spanning from 13~eV to 20~eV and centering at 16~eV.\cite{Huffman13} It forms the prompt pulse of scintillation light. There are about $1.7\times10^4$ photons per decay event at the end point energy of 782~keV in the neutron beta decay spectrum. On the other hand, the neutron capture is purely a nuclear recoil event, the scintillation process of which is similar but of different features. About 13\% of the recoil energy of 764~keV converts into a prompt light pulse and results in about $6.4\times10^3$ photons per capture event.\cite{Ito12} The stopping power $dE/dx$ for a recoiling nucleus in liquid helium of a density $\rho =$ 0.145 g cm$^{-3}$ is $2\times10^4$~eV $\mu$m$^{-1}$. The typical stopping range for a 800~keV recoiling nucleus is 40~$\mu$m. By contrast, the stopping power for a 800~keV recoiling electron is only 40~eV $\mu$m$^{-1}$\cite{NIST_db124} on average, and its stopping range can reach up to 2~cm. Therefore, a diameter of 7.5~cm is sufficient to prevent almost all of the recoiling electron born in the 3~cm diameter CN beam from touching the inner surface of the decay volume. It nearly guarantees no quenching of the prompt scintillation on the wall. Because of the dramatic difference in track length, the scintillation light of decay events is much more dispersed spatially than that of capture events. The former appears as a line of chained point sources, whereas the latter as a single point source. 

In addition, both electron and nuclear recoils also generate a large amount of triplet He$_2^*$ excimers ($a^3\Sigma_u^+$), which has a 13~s lifetime in liquid helium. The radiative decay of the triplet excimers is forbidden as it requires a spin flip; yet it can occur via the bimolecular Penning ionization that converts a portion of the triplet into singlet, most likely in a high density of triplet excimers along the recoil track. This type of scintillation light appears as
 {\color{black} a large number of after-pulses of EUV photons, following the prompt pulse, and temporally scattered over tens of micro-seconds. Each of them is much weaker than the prompt pulse, and therefore mostly registered as pulses of single or a few photo-electrons in the same detector. The occurrence rate of after-pulses} 
decreases as to a combination of two components dependent exponentially and inversely on time, respectively, $g(t) = Ae^{-t/\tau_s}+B/t+C$. It has been experimentally demonstrated that the $1/t$ component of the electron recoils is much weaker than that of the nuclear recoils.\cite{McKinsey03,McKinsey_thesis} This feature offers an important tool to distinguish the decay events from the capture events.

\section{Detection of Scintillation}
\label{sec:dection_Scint}
A standard method has been well developed to detect the EUV scintillation in liquid helium by many experiments.\cite{Huffman13, Ito12, McKinsey03, McKinsey_thesis, McKinsey04, Guo12} Based on the known technologies, we describe a conceptual design as a baseline for a quantitative analysis. A schematic of the detector is shown in Fig. \ref{fig:fig_schematic_detector}. The EUV scintillation light is first converted into a blue spectrum near 400~nm by an organic fluor -- tetraphenyl butadiene (TBP). A thin layer of evaporated TPB (eTPB) can be coated on an acrylic film and wrapped into a cylinder as the boundary of the decay volume. The eTPB coating faces the inside of the decay volume. Optical fibers can be molded with a structural support as if wound on the outside of the film cylinder to collect light. The fibers cover the full length of the decay volume so as to maximize the light collection. Wavelength shifting (WLS) fibers are a common option to convert the emitted blue light into a green spectrum near 500~nm along with a redistribution of photon phase space. A portion of the shifted light can be trapped inside the fiber by total internal reflection and transmitted to the photon sensors. 

The overall light conversion efficiency is estimated as follows. Owing to the large $L/D$ ratio of the decay volume, more than 96\% in solid angle of the scintillation light can be converted by TPB for the events occurring in the central region, as shown in Fig. \ref{fig:fig_solidAngle}. The conversion efficiency of eTPB has been demonstrated to be greater than unity.\cite{McKinsey97} Since a thick eTPB coating often appear opaque for visible light due to its surface roughness, the blue photons heading inwards the decay volume might reflect and diffuse on the eTPB coating.  
{\color{black} It is thus difficult to characterize the distribution of these inwards-going photons that are collected by WLS fibers upon multiple scattering in the eTPB coating}. 
As a moderate estimate, we only take into account the 50\% of eTPB re-emitted blue photons that travel outwards to the adjacent WLS fibers. Approximately 90\% of them can impinge on the fiber cores with the help of a 
{\color{black} Polytetrafluoroethylene (PTFE)} reflector, which is also a structural holder clamped on the outside, and then about 80\% is absorbed and shifted into green light. The double cladding WLS fibers made by Kuraray have a trapping efficiency of 5.4\% in one direction.\cite{fiber} When read on both ends, 10.8\% of the shifted green light can be conveyed towards the 2 photon sensors. Since the fiber has a bending loss of about 4\% per turn on a 7.5~cm  diameter curve and an attenuation length longer than 7.5~m, the fiber length must be constrained. In each detector unit, a round WLS fiber of 1~m long and 1~mm diameter is 
{\color{black} helically} wound around the decay volume by 3 turns, and {\color{black} the extra length on each free end is routed to a separate photon sensor.}
It needs 250 units in a tight packing to cover the whole length of the decay volume and set up an axial resolution. The average transmission efficiency of such a configuration is about 90\% along the fiber. With regard to the difficulty of making large amount of superfluid-leak-tight fiber feedthroughs, there must be two optical breaks at the windows of the liquid helium vessel, each of which has a 90\% transmission. 
{\color{black} As for the 500 photon sensors, we may employ silicon photomultipliers, which are compact in size and }have a typical quantum efficiency of 34\% for the versions with large microcells \cite{SiPM}. 
The overall conversion efficiency $\eta_{\scriptscriptstyle tot}$ is 0.9\%, {\it i.e.} an average of 9 photo-electrons ({\it PE}) can be detected per $1\times10^3$ EUV scintillation photons. The average prompt {\it PE} numbers for decay events with the spectrum peak energy at 245~keV and the endpoint energy at 782~keV are $\overline{N}_{\scriptscriptstyle PE}^{(peak)}=49.8$ and $\overline{N}_{\scriptscriptstyle PE}^{(endpt)}=158.9$, \iffalse {\color{blue} (\missingcommand{N_PE_e})}\fi respectively. 
Every {\it PE} corresponds to about 5~keV \iffalse {\color{blue} (\missingcommand{dE})}\fi of electron recoil energy. On the other hand, the average prompt {\it PE} number for the neutron capture events of recoil energy at 764~keV is $\overline{N}_{\scriptscriptstyle PE}^{(p+t)}=57.7$. \iffalse {\color{blue} (\missingcommand{N_PE_n})} \fi It coincides with beta events of 283.8~keV, \iffalse {\color{blue} (\missingcommand{cid_KE_e})} \fi close to the peak of the beta spectrum. 

% Section of Detector Response
\section{Detector Response and Event Reconstruction}

We perform a preliminary study on the response of detectors by Monte Carlo simulations. As listed below, several assumptions have been adopted to simplify the model but present the essential physics as a proof of principle. Further modelling with more details is needed.

\renewcommand{\labelenumi}{(\roman{enumi}).}
\begin{enumerate}
  \item Only the prompt scintillation signals are recorded for all the events. It means the decay and capture events cannot be distinguished among the simulated data. Yet in real experiment, they are distinguishable by the difference in the $1/t$ time-dependent occurrence rates of the after-pulses. {\color{black} This additional information will improve data analysis and understanding of systematic effects} 
  \item The scintillation light for the capture events is emitted from a point source as their track length is tens of microns, whereas that for the beta events is from an energy-dependent straight tracks of length up to 2~cm. For electron recoils, more energy deposits in the vicinity of the track end as it slows down. The spatial energy deposition approximately follows $dE/dr \propto r^2$, where $r$ is the geometric distance from the starting point of electron recoil.\cite{Adams01} Further studies can be performed on simulated scattering tracks with productions of secondary $\delta$-electrons. 
  \item Only the outward-going portion of the eTPB converted light can be collected by the adjacent fibers, but none of the inward-going, as the latter reflects and diffuses on the coating into a broader distribution over all the detectors, yet much weaker in intensity than the former. 
  \item The EUV light converted by eTPB will be collected by the fibers tightly wound against the thin film at the same axial position. It means the solid angle of light from an event projecting on the section of the eTPB film is equivalent to that on the detector lying against the film. 
  \item All the detectors have the same efficiency. In reality, the efficiency of detectors are different and may vary with time. Calibrations are necessary and will be discussed in Subsection \ref{subsec:Calibration}.
  \item There is no timing information in this simulation. We assume all the events are in the coincident time window as the cold neutron beam passes the decay volume. Timing information will greatly improve the position reconstruction and be a practical way of minimizing background.
\end{enumerate}

\begin{figure}
  \centering
  \includegraphics[width=5.5in]{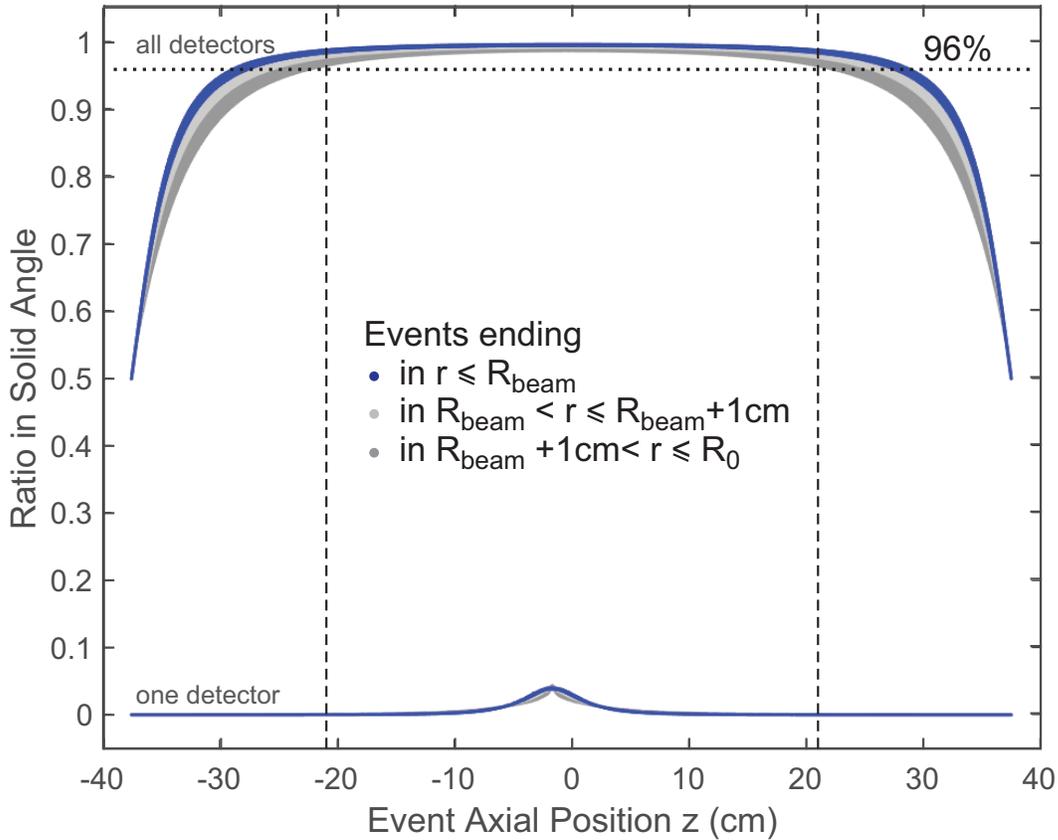}
  \caption{\label{fig:fig_solidAngle} The coverage of solid angle for events ending at different axial positions in the decay volume. The upper plot is the total coverage of solid angle by the sum of all the detectors; and the lower plot is that by one unit detector centered at $-1.65$~cm. Events ending in the beam pass (r $\leqslant$ R$_{\mbox{beam}}$) are marked with blue, within 1cm away from the beam pass (R$_{\mbox{beam}}<$ r $\leqslant$ R$_{\mbox{beam}}$+1~cm) marked with light grey, and outside the regions above (R$_{\mbox{beam}}$+1~cm $<$ r $\leqslant$ R$_{\mbox{0}}$) marked with dark grey, where R$_{\mbox{0}}$ is the maximum radius that a recoiling electron can reach. }
\end{figure}

\begin{figure}
  \centering
  \includegraphics[width=5.5in]{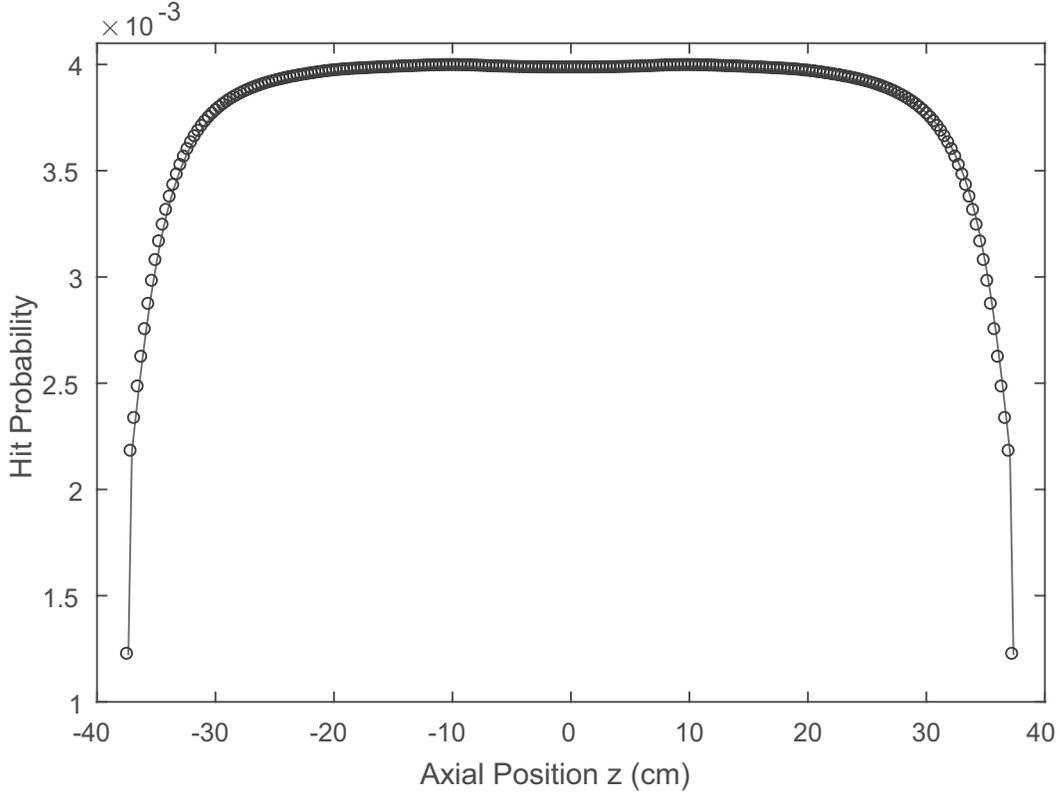}
  \caption{\label{fig:fig_det_prob} The overall hit probability on each of the 250 detectors}
\end{figure}

A sampling on starting position $\overrightharp{x}_0$ of events as to a uniform distribution function, $Pr(\overrightharp{x}_0)=const.$, is carried out in the beam-occupied volume.\footnote{The cross sectional distribution of the beam fluent rate $I(v,\overrightharp{r})$ is independent of $\overrightharp{r}$ in this simulation.} $4\times10^7$ random events are generated and assigned as either capture or decay according to a preset ratio, $\kappa=4.4975$. The capture events are of a point source in the beam-occupied volume, whereas the decay events are of a straight line source that may extend out. For each event, emission of scintillation light is isotropic, and the portion of solid angle received by each and every of the 250 detectors is simulated as the hit probability of each detector given the source position $\overrightharp{x}$,  $Pr(n_{det} | \overrightharp{x})$, where $n_{det}\in[1,250]$ is the detector number. Fig. \ref{fig:fig_solidAngle} plots the accepted portion of solid angle by one unit detector centered at $-1.65$~cm, and sum of that over all the detectors for each event at various axial positions $z$, respectively. The events ending in the beam path are highlighted in blue. It shows the total accepted solid angle has a weak dependence on the radial ending position, which correlates with the recoil energy of electrons. With all the detectors functioning, it can cover more than 96\% of solid angle for events in a central region spanning 42~cm in the axial direction, as shown by the upper plot in Fig. \ref{fig:fig_solidAngle}. The events lying within the axial edges of a unit detector have about 4\% of chance to be registered by this detector, as shown in the lower plot in Fig. \ref{fig:fig_solidAngle}. The overall detector hit probability $Pr(n_{det})$ is derived by the integral over the entire volume $V_0$, $Pr(n_{det}) = \int_{V_0} d\overrightharp{x}~Pr(n_{det}|\overrightharp{x})~ Pr(\overrightharp{x})$, and plotted in the Fig. \ref{fig:fig_det_prob}. For detectors in the central region, the overall hit probability is about 0.40\% for scintillation light of an event at any position to be registered; while for those close to the ends, the chance is naturally much less. 

\subsection{Reconstruction of Event Position and the Fiducial Cut}
\label{subsec:ReconsPosition}
\begin{figure}
  \centering
  \includegraphics[width=5.5in]{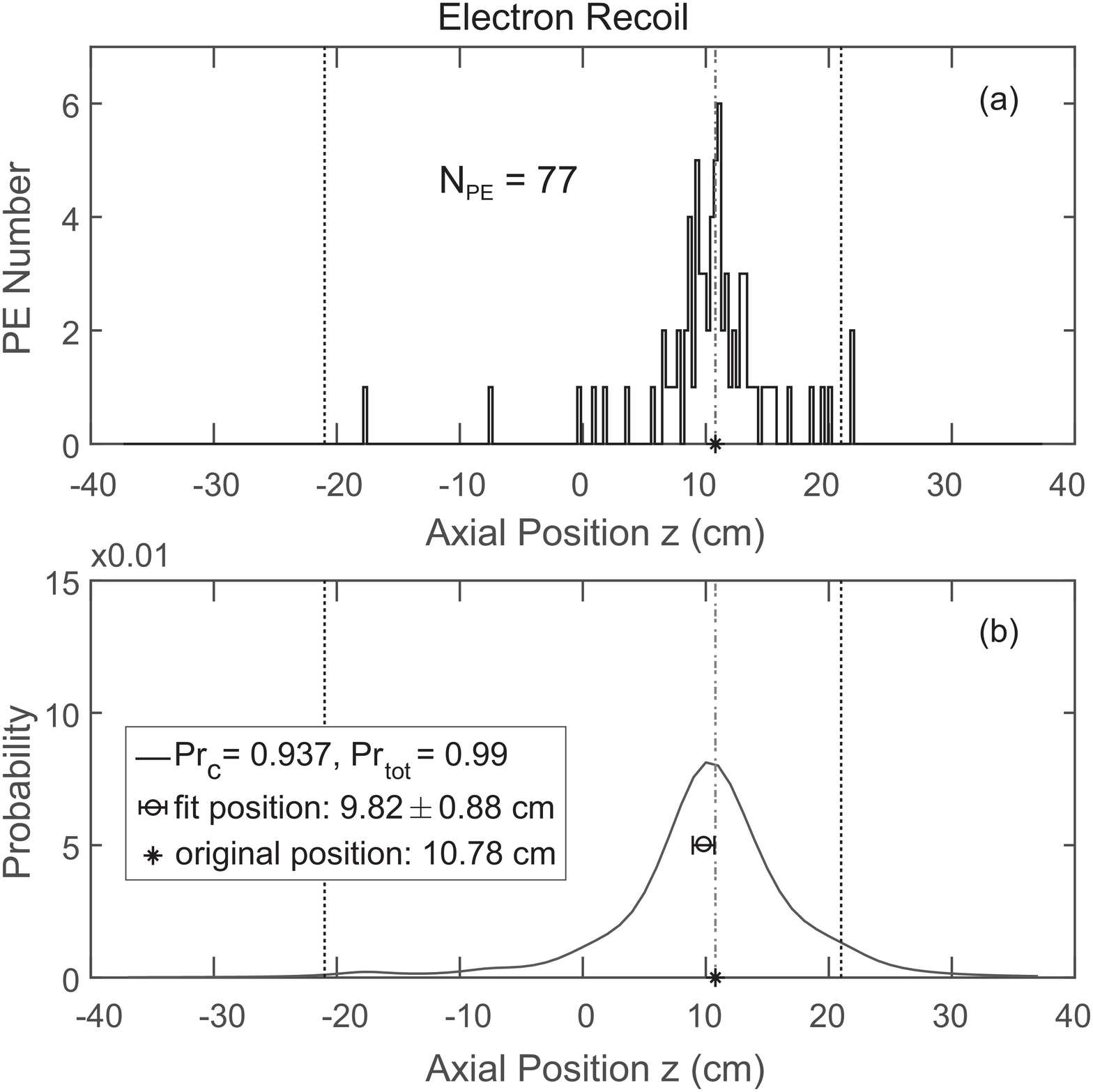}
  \caption{An example on the deduced possibility of event axial position based on the observed distribution of 77 {\it PE}\hspace{0.1em}s. 
  (a) the spatial distribution of observed {\it PE} numbers on different detectors; (b) the spatial distribution of probabilities on different event axial positions. The total probability Pr$_{tot}$ is 0.99 and the probability in the central region Pr$_{c}$ is 0.937. The fit position based on the observed {\it PE}\hspace{0.1em}s is $9.82\pm0.88$~cm, and in comparison, the original position fed in the simulation is 10.78~cm. }
  \label{fig:fig_locationProb}
\end{figure}

\begin{figure}
  \centering
  \includegraphics[width=5.5in]{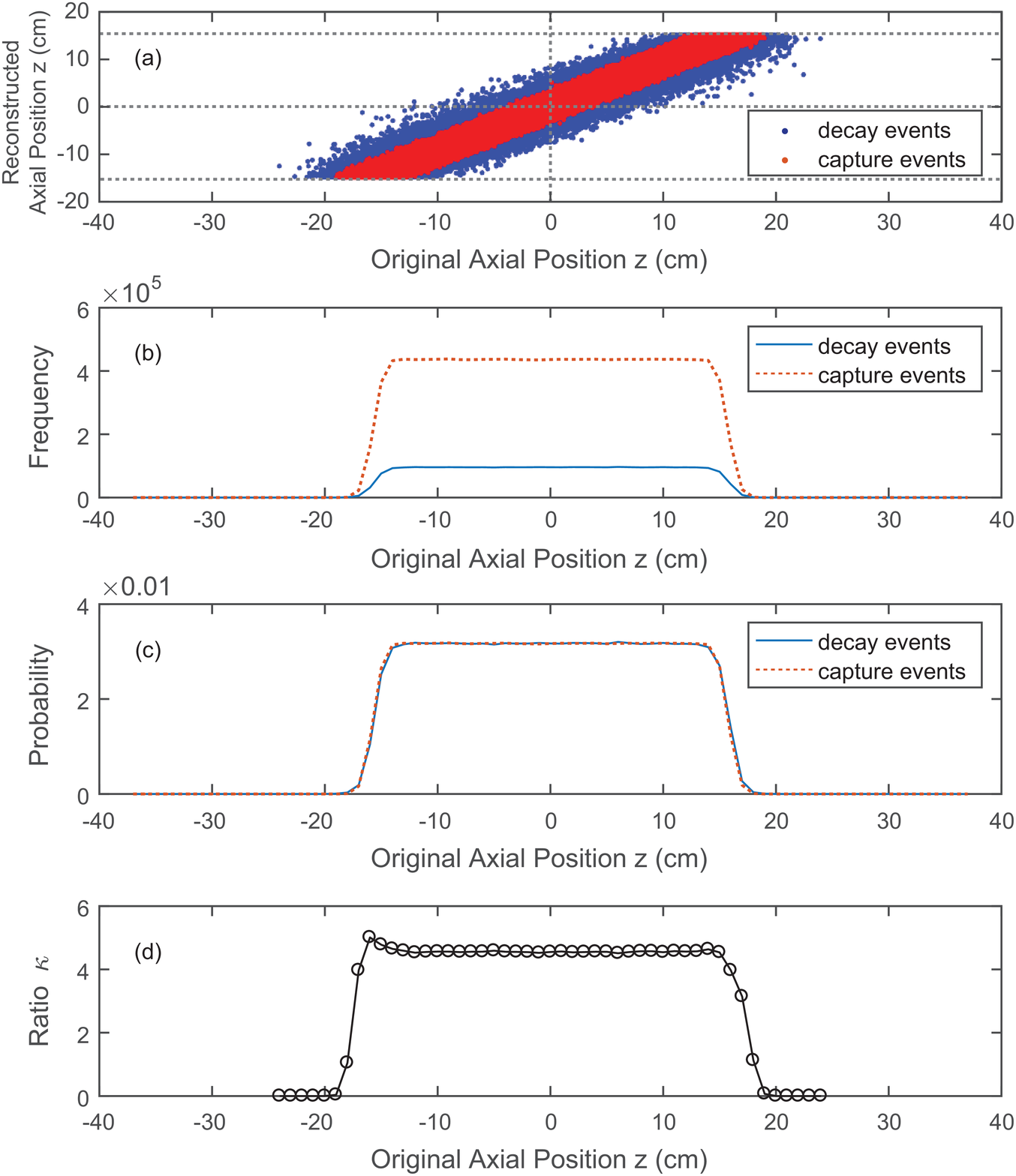}
  \caption{Analysis of the combined fiducial cut on the simulated data. (a) the reconstructed axial position $z$ of both the simulated decay events (blue dots) and capture events (red dots) plotted against their original position; (b) and (c) frequency and probability distribution of the selected decay events (blue solid line) and capture events (red dashed line) by the fiducial cut as to their original positions, respectively; (d) the ratio $\kappa$ calculated within each bin of original position $z$. }
  % NOTE: the legend for fig1 is opposite.
  \label{fig:fig_HistPositionRecon}
\end{figure}

With the registered {\it PE} numbers $N_{\scriptscriptstyle PE}(n_{det})$ from an event on a series of detectors $n_{det}, 1 \leqslant n_{det} \leqslant 250$, the probability of event position within a given sub-volume $\Delta V(z_k)$ centered at $z_k$ is derived in Eqn. (\ref{eqn:event_position_prob}) by the Bayes' theorem. It is on the assumption that every detector is independent, {\it i.e. }no cross talk. The probability of an event within a certain region of interest, {\it e.g.} the 42~cm long central region, can be calculated as the accumulated probability, $\sum_{z_k}Pr(\Delta V(z_k)|N_{\scriptscriptstyle PE}(n_{det}))$. Fig. \ref{fig:fig_locationProb}a shows an example of the simulated electron recoil event with 77 observed {\it PE}\hspace{0.1em}s distributed on several detectors. The entire volume $V_0$ is divided into sub-volumes $\Delta V(z_k)$ as disks of 1~cm thick, and the distribution probability of event position for each sub-volume is calculated and plotted in Fig. \ref{fig:fig_locationProb}b. The accumulated probability for this event to occur within the 42~cm long central region is 93.7\%. In general, the more {\it PEs} observed, the more accurate the reconstructed position of the event. However, since the events near the ends of the decay volume lose a significant portion of scintillation light on the end windows, {\it i.e.} information is truncated, the reconstructed positions are biased towards the center. Therefore, the accumulated probability of positions inside the central region of 42~cm generally performs better in identifying events thereof than that of reconstructed positions, especially for low {\it PE} events. 
Yet neither is satisfactory in selecting events in a region with relatively identical position distribution and uniform ratio of capture-to-decay event rates. Later, we find a combination of both actually forms a good fiducial cut. It is demonstrated in Fig. \ref{fig:fig_HistPositionRecon}. Figs. \ref{fig:fig_HistPositionRecon}a and \ref{fig:fig_HistPositionRecon}b show the events are selected by a combination of the following criteria, (i). reconstructed positions within $\pm15$~cm, and (ii). accumulated probability of more than 80\% inside the central region of 42~cm. The number of selected events is about 42\% of the total. 
Fig.\ref{fig:fig_HistPositionRecon}c shows the distributions of the original axial positions between the decay and capture events are almost identical; and Fig.\ref{fig:fig_HistPositionRecon}d shows the capture-to-decay ratio $\kappa$ has a flat plateau and relatively sharp edges in the selected region with respect to the original axial positions.

\begin{equation} 
\label{eqn:event_position_prob}
    \begin{split}
         Pr(\Delta V(z_k)|N_{\scriptscriptstyle PE}(n_{det}), & 1 \leqslant n_{det} \leqslant 250) \\ 
         = & \int_{\Delta V(z_k)} d\overrightharp{x} ~ \frac{ \sum_{n_{det}=1}^{250} N_{\scriptscriptstyle PE}(n_{det}) ~ Pr(n_{det}|\overrightharp{x}) ~ Pr(\overrightharp{x})  }{\sum_{n_{det}=1}^{250} N_{\scriptscriptstyle PE}(n_{det}) ~ Pr(n_{det})} 
    \end{split}
\end{equation}

\subsection{Determination on the Ratio of Capture-to-Decay Event Rates}
\label{subsec:ratio_capt_decay}
\begin{figure}
  \centering
  \includegraphics[width=5.5in]{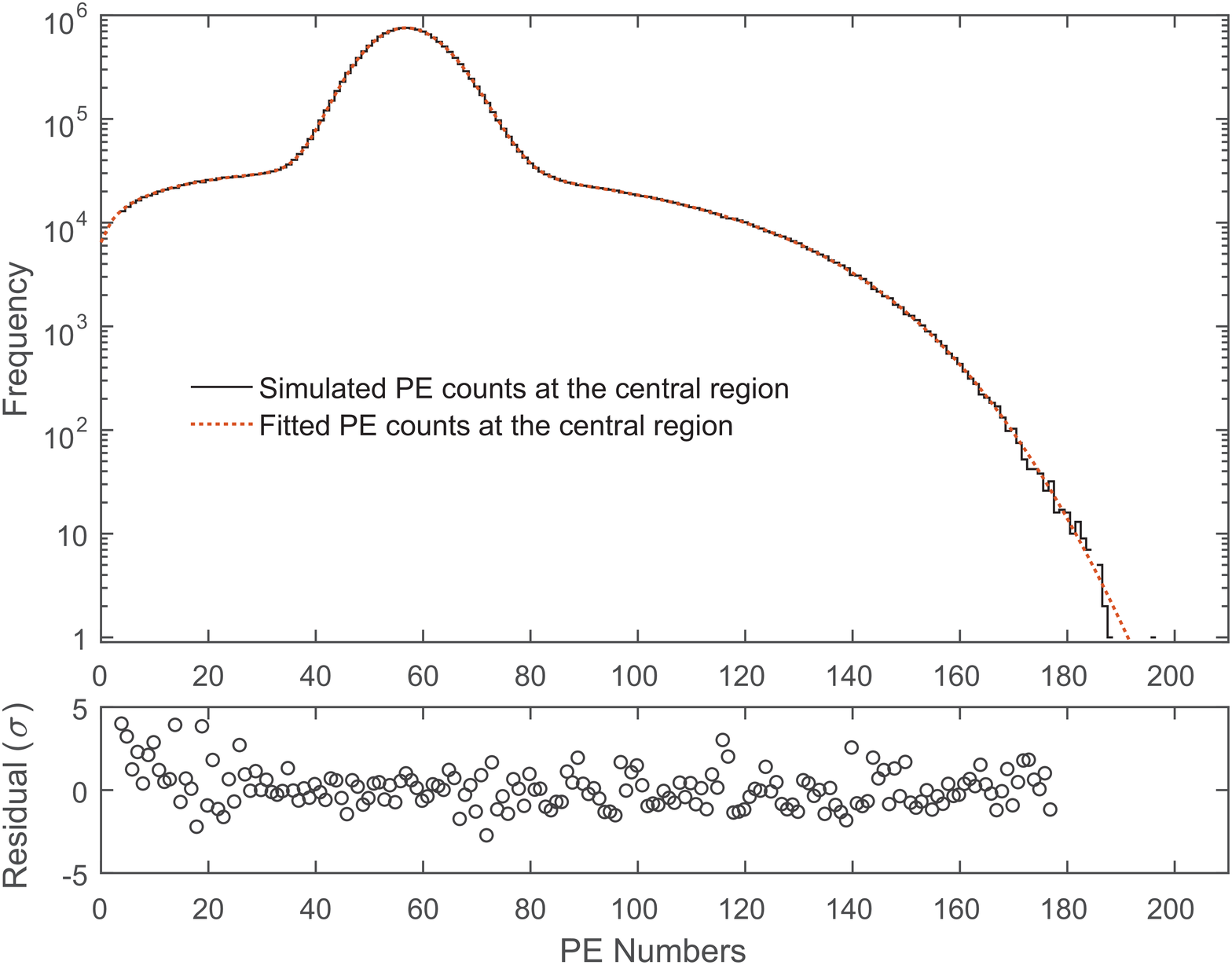}
  \caption{Simulated neutron beta decay spectrum in addition to the neutron capture peak (black solid line) and the result of the ML fit (red dotted line). The residual of fitting with respect to the {\it PE} numbers is shown in the lower plot. $\chi^2/ndf=1.4$}
  \label{fig:fig_FitSpectrum}
\end{figure}

The spectrum of the selected events is plotted against the {\it PE} number in Fig. \ref{fig:fig_FitSpectrum}. The capture events overlap with the decay events in the mid range and the spectrum cuts off at a lower bound of 4~{\it PE}\hspace{0.1em}s, equivalent to a recoil energy {\it KE}$_e=$ 20~keV, on purpose to exclude random backgrounds of few photons. In order to resolve the ratio $\kappa$ of capture-to-decay event rates from the acquired spectrum, a theoretical model for fitting is constructed. It consists of three components: the neutron decay spectrum, the single capture peak and the background. In this study, we only simulate signals of the former two, but omit the effect of the background, because it will be poorly defined without the knowledge of the actual system. A discussion on the possible backgrounds will be presented in Section \ref{sec:Background}. 

\begin{figure}
  \centering
  \includegraphics[width=5.5in]{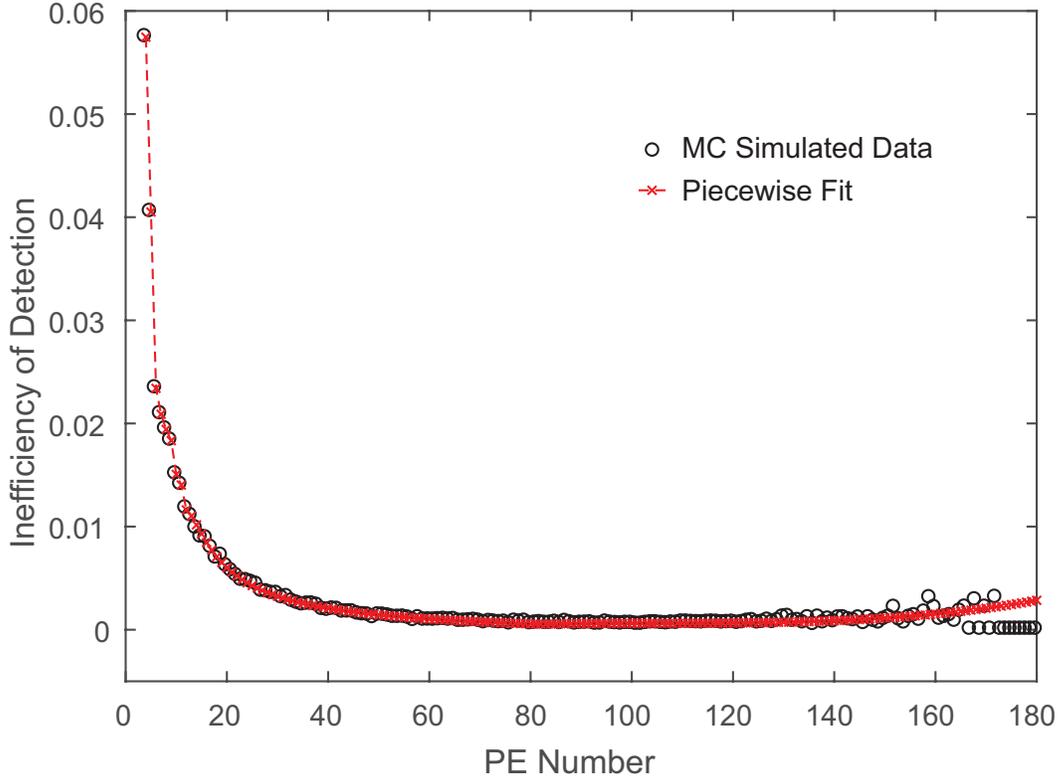}
  \caption{Inefficiency of detection, $1-\epsilon_{\scriptscriptstyle cut}(N_{\scriptscriptstyle PE})$, related to the combined fiducial cut of reconstructed positions within $\pm15$~cm and 80\% accumulated probability within $\pm21$~cm. The simulation data is plotted in black circles, and the piecewise fit is plotted in red dashed line. 
  {\color{black} Piecewise fit can better capture the features of inefficiency as a sharp rise in low {\it PE} region and a slow tilt in the high {\it PE} region.} 
  }
  \label{fig:fig_cut_ineff}
\end{figure}

The neutron beta decay spectrum is formulated as 

\begin{equation}
    \label{eqn:neutronBetaDecayFunction}
    \frac{d\Gamma_n}{dE_e} \propto F_n(E_e) p E_e (E^{(endpt)}-E_e)^2
\end{equation}

where $E^{(endpt)}$ is the endpoint energy, $E_e$ is the total electron energy, and $p=(E_e^2-m_e^2)^{1/2}$ is the momentum of electron. $m_e$ is the electron mass. $F_n(E_e)$ is the Fermi function for neutrons defined as 

\begin{equation}
    \label{eqn:neutronFerminFunction}
    F_n(E_e)=4 \exp{(\pi \alpha E_e/p)} \frac{|\Gamma(1-i\alpha E_e/p)|^2}{\Gamma(3)^2}
\end{equation} 

where $\alpha=1/137$ is the fine structure constant, and $\Gamma(z)$ is the gamma function. In reality, the spectrum contains a Poisson smearing due to the random process in the light transportation and conversion into {\it PE}\hspace{0.1em}s. Though the deposited energy from neutron capture events is single-valued at 764~keV, it appears as a much broadened peak. 
The Poisson probability function, $Pois\left[N_{\scriptscriptstyle PE}, \lambda\right]$, with the {\it PE} number $N_{\scriptscriptstyle PE}$ and the mean $\lambda$, is thus embedded in the probability function $f(N_{\scriptscriptstyle PE})$ of the observed spectrum as  

\begin{equation}
\label{eqn:fitTheoreticalSpectrum}
    \begin{split}
        f(N_{\scriptscriptstyle PE}) = &  \frac{1}{1+\Tilde{\kappa}}~\frac{\epsilon_{\scriptscriptstyle cut}(N_{\scriptscriptstyle PE})}{C_e}~\sum_{M_{\scriptscriptstyle PE}}  Pois\left[ N_{\scriptscriptstyle PE},M_{\scriptscriptstyle PE}\right] \cdot \mathcal{F}_e(M_{\scriptscriptstyle PE})
        \\
        & + \frac{\Tilde{\kappa}}{1+\Tilde{\kappa}}~\frac{1}{C_{p+t}}~ Pois\left[N_{\scriptscriptstyle PE}, \lambda_{p+t}\right] + B 
    \end{split}
\end{equation}

where $\Tilde{\kappa}$ is the observed ratio of capture-to-decay event rates, $C_e$ and $C_{p+t}$ are the normalization factors for the spectrum bins above the lower bound of 4 {\it PE}\hspace{0.1em}s. The third term $B$ represents the background. The second term is the broaden peak of the neutron capture events, where $\lambda_{p+t}$ is the mean {\it PE} number for the capture events. The first term is the modified neutron beta decay spectrum. In a Poisson process, events of different recoil energies and their corresponding deterministically-converted {\it PE} number $M_{\scriptscriptstyle PE}$ all contribute to a given bin of observed {\it PE} number $N_{\scriptscriptstyle PE}$. The share of contribution from each $M_{\scriptscriptstyle PE}$ is actually the probability of events with a deterministic $M_{\scriptscriptstyle PE}$ in the neutron beta decay spectrum that follows Eqn. (\ref{eqn:neutronBetaDecayFunction}), namely $\mathcal{F}_e(M_{\scriptscriptstyle PE})$, and given by

\begin{equation}
        \mathcal{F}_e(M_{\scriptscriptstyle PE}) = \int_{(M_{\scriptscriptstyle PE}-0.5)/\eta_e\Omega(M_{\scriptscriptstyle PE})}
        ^{(M_{\scriptscriptstyle PE}+0.5)/\eta_e\Omega(M_{\scriptscriptstyle PE})} ~\frac{d\Gamma_n}{dE_e}~dE_e 
\end{equation}

where $\eta_e\cdot\Omega(M_{\scriptscriptstyle PE})$ is the  {\it KE}$_e$-to-{\it PE}\hspace{0.1em}s conversion coefficient for electron recoils. As indicated in Fig. \ref{fig:fig_solidAngle}, the coverage ratio of solid angle $\Omega(M_{\scriptscriptstyle PE})$ has a weak dependence on the electron recoil energy, {\it i.e.}, on the {\it PE} number $M_{\scriptscriptstyle PE}$ in the central region of the decay volume. The particular coefficient $\epsilon_{\scriptscriptstyle cut}(N_{\scriptscriptstyle PE})$, in the first term of Eqn. (\ref{eqn:fitTheoreticalSpectrum}),  is the efficiency of detection related to the fiducial cut. It comes from a significant position uncertainty due to few {\it PE}\hspace{0.1em}s and long tracks for the low and high {\it PE} events, respectively. It inevitably introduces a detection inefficiency, $1 - \epsilon_{\scriptscriptstyle cut}(N_{\scriptscriptstyle PE})$, and hence, a distortion of spectrum. Such an effect is extracted from the simulation and fitted as shown in Fig. \ref{fig:fig_cut_ineff}. 

There remain 3 fitting parameters in Eqn. (\ref{eqn:fitTheoreticalSpectrum}): $\eta_e$, $\lambda_{p+t}$, and most importantly, $\Tilde{\kappa}$. Maximum Likelihood (ML) method is employed to acquire the best fit and associated errors of the above 3 parameters. $\kappa$ in Eqn. (\ref{eqn:fitTheoreticalSpectrum}) is marked as $\Tilde{\kappa}$, as the fitting only acquires the observed capture-to-decay ratio within the data bins. $\Tilde{\kappa}$ must be corrected for the actual $\kappa$ with the detection efficiencies $\epsilon_{\scriptscriptstyle He3}=1$ and $\epsilon_{\beta}=0.9878$, where $\epsilon_{\beta}<1$ is due to the missing portion of spectrum less than 4 {\it PE}\hspace{0.1em}s. The result is listed in Table \ref{tab:MLFitResult}, and the plots in Fig. \ref{fig:fig_FitSpectrum} show the result of the ML fit (in red dotted line) on top of the acquired spectrum (in black solid line). It demonstrates the fit can extract $\kappa$ value at an accuracy well within 0.1\%. 

\begin{table}
\lineup
\caption{\label{tab:MLFitResult}Result of ML fit to the simulated spectrum.}
\begin{indented}
    \item[]\begin{tabular}{lcccc}
        \br
        & $\kappa=\Delta N_{p+t}/\Delta N_{\beta}$  & $\eta_e$ & $\lambda_{p+t}$ \\
        \mr
        Preset Value     & 4.4975 & 0.2032  & 57.254  \\
        Monte Carlo & 4.4997 & 0.2032 & 57.292  \\
        ML Fit \iffalse{($\chi^2/ndf$=1.4)}\fi \iffalse (C.L.>0.96)\fi & $4.4983\pm0.0040$  & $0.2037\pm0.0001$ & $57.291\pm0.002$ \\
        \br
    \end{tabular}
\end{indented}
\end{table}

\subsection{Reconstruction of Event Energy and Detector Calibration}
\label{subsec:Calibration}
Reconstruction of event energy will be performed in calibration of all the detectors as a cross reference. In the previous subsection, the conversion efficiency $\eta_e$ for electron recoils is fitted through the analysis of the neutron decay spectrum, and the reconstructed energy is obtained as $E_e = N_{\scriptscriptstyle PE}/\eta_e\Omega(N_{\scriptscriptstyle PE})$. In reality, each photon sensor has a different quantum efficiency $\eta_{\scriptscriptstyle SiPM}^{(i)}$, and each fiber has a variation in transmission efficiency $\eta_{\scriptscriptstyle fiber}^{(i)}$. The overall detection coefficient $\eta_{\scriptscriptstyle tot}(\overrightharp{x})$ thus varies for different sub-volumes, due to the variation of solid angles upon detectors of different quantum efficiency. It is a common approach to use conversion electron sources, such as $^{109}$Cd (63, 84~keV), $^{139}$Ce (127, 160~keV), $^{113}$Sn (364, 388~keV), $^{207}$Bi (481, 975, 1047~keV) for calibrations in between production runs.\cite{Plaster12} These sources can be placed in many designated positions to map out the response of different detectors. During the production runs, the calibration can also be done with the neutron capture peak, and additional deposits of $\alpha$ source or lithium neutron capture film on the end windows of the decay volume.

\section{Background Suppression}
\label{sec:Background}

In order to achieve a highly accurate measurement on the ratio of capture-to-decay rates, background signals must be properly suppressed, discriminated or subtracted. Cosmic ray muons can be easily identified by coincidence in the veto detectors surrounding the apparatus. The static radioactive backgrounds from materials of the apparatus can be shielded by a thick layer of lead or tungsten as shown in Fig. \ref{fig:fig_schematic_detector} and characterized in the background runs. The gamma rays from the cold neutron source can be greatly suppressed by bending the beam direction out of sight with proper neutron optics.\cite{Fomin15} The most harmful type of backgrounds are the gamma rays produced by the neutron-induced activation near the decay volume and undergoing Compton scattering on liquid helium inside the decay volume and the surrounding WLS fibers. The Compton electrons are identical to the decay events on the features of scintillation signals. It hence will be pooled in the {\it PE} spectrum and modelled as $B$ in Eqn. (\ref{eqn:fitTheoreticalSpectrum}). The spectrum of the Compton electrons produced by gamma rays above 4~MeV is mostly flat in the region of neutron decay spectrum. Some of the delayed gamma rays can be characterized during the intervals between the CN beam pulses, such as the 1.6~MeV gamma rays emitted at a half-life of 11.16~s from the neutron activated fluorine. In this paper, we focus on two types of prompt gamma rays due to the neutron captures by the window material, and by hydrogen, $p+n \rightarrow d + \gamma~(2.2 ~\mbox{MeV})$, in the plastic components surrounding the decay volume, such as the WLS fibers, {\it etc.} They are believed to be the major contributors to the backgrounds.

The first measure to suppress neutron-induced gamma rays is to reduce the capture and scattering of neutrons on the window materials. Polycrystalline CVD diamond is a good option, because carbon has relative small scattering and capture cross sections, and a thin window of 5~cm diameter and 1~mm thickness \cite{diamond_window} is commercially available with a good mechanical strength. The capture cross section of carbon for the 17~\AA~cold neutrons is 0.033 barns and the capture fraction is $5.66\times10^{-4}$. For a cold neutron flux of $1.5\times10^7$~CN~s$^{-1}$, \iffalse {\color{blue} (\missingcommand{CN_incident_rate})} \fi
about $8.6\times10^3$~Hz \iffalse {\color{blue} (\missingcommand{CN_win_capt_dN_dt})} \fi of neutrons are captured with an emission of prompt gamma rays mostly at energies of 1.3, 3.7 and 4.9~MeV. A simulation shows the intensive prompt gamma rays result in more than 600~Hz \iffalse {\color{blue} (\missingcommand{gamma_win_capt_LHe})} \fi Compton events in liquid helium inside the decay volume, and more than 100~Hz \iffalse {\color{blue} (\missingcommand{gamma_win_capt_fib})} \fi in the polystyrene WLS fibers. Most events distribute spatially near the windows, and temporally at the moments when the neutron flux passes the windows. Although the number of the window-originated Compton events greatly overwhelms that of the decay events, they can be separated in time if the neutron beam can be chopped into sharp pulses both in time and energy. It requires the decay volume to be set up close to the source. The neutron decays will then appear as scattered single events in time sequence between two intensive bursts of Compton events when neutrons pass the entrance and exit windows, respectively. Considering it takes about 3.2~ms for the 17~\AA~neutrons to pass the 75~cm long decay volume, the middle 1.6~ms is the time interval when the beam pulse passes the central region. The typical recovery time for the SiPM sensors are hundreds of nano-seconds\cite{SiPM}, whereas the rate of after pulses following each prompt signal decays in tens of micro-seconds\cite{McKinsey03}. Therefore, such a time cut can effectively distinguish the events occurring in the central region, which are crucial in construction of the energy spectrum, and eliminate impacts of the window-originated Compton events.  In return, the bright bursts of Compton events can be used as a calibration reference of the beam flux and spectrum.

The scattering of cold neutrons on the windows at 0.5~K is dominated by incoherent scattering, which is an s-wave scattering independent of the incident velocity. The incoherent cross section of carbon for the 17~\AA~cold neutrons is 0.001 barns, and the scattered fraction is $1.76\times10^{-5}$. For the same cold neutron flux as above, about 270~Hz \iffalse {\color{blue} (\missingcommand{CN_win_scat_dN_dt})}\fi of neutrons are scattered isotropically from both the windows into the delay volume and interact with hydrogens on the fibers. The incoherent scattering cross section of hydrogen is 80.3 barns, much larger than the capture cross section of 3.1 barns for 17~\AA~cold neutrons. The stray neutrons will mostly scatter incoherently in the plastics and has a small chance of being captured by the hydrogen nuclei. In order to minimize the chance of captures, a second measure is to deploy an effective neutron absorber on the outside of the plastics so as to capture all the outgoing stray neutrons. Lithium-6 enriched material is an ideal option, since $^6$Li has a large neutron absorption cross section of 8887 barns for 17~\AA~cold neutrons, and there is no associated emission of gamma rays in the reaction,~~$^6Li+n \rightarrow \alpha + t + 4.78~\mbox{MeV}$. 
% $^6Li+n \rightarrow \alpha~(2.73MeV) + t~(2.05MeV)$

\begin{figure}
  \centering
  \includegraphics[width=5.5in]{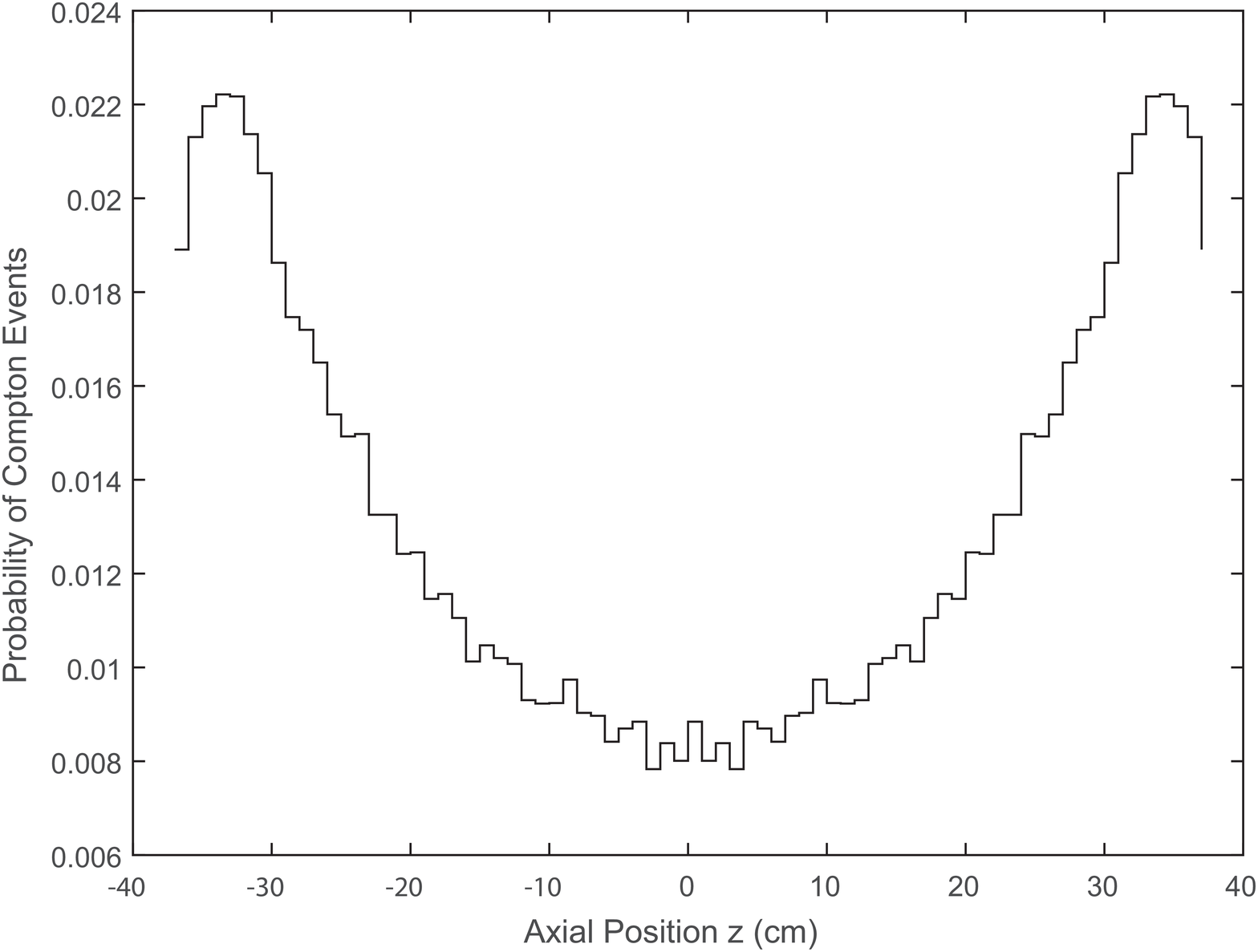}
  \caption{\label{fig:fig_zCompton} Probability of Compton electron events in the axial axis of the decay volume induced by the prompt gamma rays due to hydrogen-captures of neutrons scattered from the beam inlet and outlet diamond windows.}
\end{figure}

A simulation on neutron scattering and capture is carried out on the geometry of a 1~mm thick polystyrene fibers around the decay volume, a 5~mm thick PTFE holder clamped on the fibers, and a sufficiently thick lithium absorber at the outermost shell that absorbs all the stray neutrons, as illustrated in Fig. \ref{fig:fig_schematic_detector}. It is found about 4.3\% of the scattered neutrons are captured by the fibers and 0.4\% by the PTFE holder. The neutron captures on hydrogen in the fibers do not induce any significant scintillation in the polystyrene, as the kinetic energy of deuterium is merely about 1.3~keV. The resultant prompt gamma rays of 2.2~MeV contribute a background of Compton events at 0.28~Hz \iffalse {\color{blue} (\missingcommand{gamma_win_scat_LHe})}\fi in liquid helium inside the decay volume, and 0.31~Hz \iffalse {\color{blue} (\missingcommand{gamma_win_scat_fib})}\fi in the polystyrene WLS fibers. As shown in Fig. \ref{fig:fig_zCompton}, the Compton events have a higher chance to occur near the windows. Since the majority of Compton electrons are at energies near the Compton peak of 2.0~MeV, it adds about 0.05\% to the total counts of neutron decay events with a fiducial cut of $\pm15$~cm on the central region. It can be characterized and corrected in data analysis. 

% 16.5A CN beam of 3.1e+07 Hz scattered rate by diamond window: 5.38e+02 Hz
% Prompt Gamma Rays of 2.2 MeV turn into Compton scintillation in LHe: 0.56 Hz
% Prompt Gamma Rays of 2.2 MeV turn into Compton scintillation in fiber: 0.62 Hz

%Section of helum-3 density measurement by UCN
\section{Accurate Measurement of Helium-3 density with UCNs}

The last quantity crucial to determine the neutron lifetime in Eqn. (\ref{eqn:nTauEqu}) is the $^3$He density $n_{\scriptscriptstyle He3} = 2.18\times10^{22} X_{\scriptscriptstyle He3}$, where the fractional concentration $X_{\scriptscriptstyle He3}$ of about $2\times10^{-10}$ needs to be prepared and characterized in high precision. 
{\color{black} One possible way of measurement is to employ the Atom Trap Trace Analysis, which has been demonstrated to measure the abundance of rare isotope $^{39}$Ar at the level of $10^{-16}$.\cite{Jiang11} Similar technology may be developed for detection of the $^3$He concentration at a precision well below 0.1\%. Then, the uncertainty of the combined term $\sigma^{th}_{\scriptscriptstyle He3} v^{th}_n n_{\scriptscriptstyle He3}$ in Eqn. (\ref{eqn:nTauEqu}) will be dominated by that of the $^3$He capture cross section of thermal neutrons, $\sigma_{\scriptscriptstyle He3}^{th}=5333\pm7$~b\cite{He3Capt}, equivalent to 0.13\%. However, in this paper, we provide another possible method to measure the $^3$He density, which takes advantage of the scintillation rates correlated to the neutron captures on $^3$He nuclei in the sample liquid helium.}
These deployed neutrons are not cold neutrons in a beam but rather ultracold neutrons (UCN) stored in a material bottle filled with the sample liquid helium. Since UCNs uniformly distribute in the storage volume and scintillations can partially quench on the walls, it is impractical to construct a well-defined spectrum as in Subsection \ref{subsec:ratio_capt_decay}, 
{\color{black} and therefore, the difference in the $1/t$ responses of after-pulses will be the key tool to distinguish the decay and capture events.} 
We will characterize the purity of $^4$He and accurately measure $^3$He concentration in the later prepared helium mixture via the time-dependent rates of both decay and capture scintillation events. 
{\color{black}Furthermore, it will be shown that value of the combined term $\sigma^{th}_{\scriptscriptstyle He3} v^{th}_n n_{\scriptscriptstyle He3}$ will be directly obtained in experiment. Therefore, our goal sensitivity in neutron lifetime of below 0.1\% can be achieved with enough statistics, and is no longer limited by the uncertainty of 0.13\% in $\sigma_{\scriptscriptstyle He3}^{th}$. }Details will be articulated in the following subsections.

\subsection{UCN Storage in a Neutron-Friendly Volume}
\label{subsec:ucn_storage}

Suppose the sample liquid helium fills a storage volume made of UV transmitting acrylic tube of 75~cm long, 7~cm ID and 7.5~cm OD. It is sealed at both ends and coated with deuterated films on the inside so that it is hermetic and friendly to UCNs. It can be installed inside the detector setup as described in Section \ref{sec:dection_Scint}, except the PTFE reflector can be as thin as 50~$\mu$m because of the following two reasons: (i). the structural support can be loaded to the storage tube; and (ii). PTFE generates a high level of background due to the delayed gamma rays from neutron-activated fluorine as discussed in Subsection \ref{subsec:UCN_gamma}. 

The UCN storage follows 
\begin{equation}
    \dot{N}_{UCN}(t) =-\frac{N_{UCN}(t)}{\tau_{tot}}, 
    \label{overview9} 
\end{equation}

\begin{equation}
    N_{UCN}\left(  t\right) = N_0 \exp{\left(- \frac{t}{\tau_{tot}} \right)},
    \label{overview10}
\end{equation}

where $N_0$ is the initial number of UCNs and $\tau_{tot}$ is the storage time constant. Several factors contribute to $\tau_{tot}$ of this volume, 

\begin{equation} \label{eqn:tau_tot_UCN}
    \tau_{tot}^{-1} = \tau_{\scriptscriptstyle He3}^{-1}+\tau_{\beta}^{-1}+\tau_{up}^{-1}+\tau_{loss}^{-1}.
\end{equation}

$\tau_{\scriptscriptstyle He3}$ is the neutron $^3$He capture time constant of interest as given in Eqn. (\ref{eqn:tau3_UCN}). For $X_{\scriptscriptstyle He3}=2\times10^{-10}$, $\tau_{\scriptscriptstyle He3}$ is about 195.2~s, \iffalse {\color{blue} (\missingcommand{tau_3})}\fi which dominates the total storage time, compared to the neutron lifetime $\tau_{\beta}\approx$880~s.

% This equation might need a He3 and neutron spectrum modification, to derive from Boltzmann distribution
\begin{equation} \label{eqn:tau3_UCN}
    \frac{1}{\tau_{\scriptscriptstyle He3}} = n_{\scriptscriptstyle He3} \sigma_{\scriptscriptstyle He3}^{th}v_{\scriptscriptstyle He3}^{th} = 2.56\times
10^{7}X_{\scriptscriptstyle He3}~[\mbox{s}^{-1}] 
\end{equation}

UCNs suffer a loss from captures or up-scattering of the wall nuclei. Such an effect can be described by Schr\"odinger equations with one-dimensional potential and characterized by a loss probability per bounce, $f(E_{\scriptscriptstyle UCN})$. The rate of wall collisions is given by $\left( vA/4V\right)$, where $v$ is the UCN velocity, $A$ is the area of the storage chamber, and $V$ is its volume. The contribution of wall losses to the storage time is then given as 
% This equation might need a derivation for a cylindrical cell?
\begin{equation}
    \frac{1}{\tau_{wall}} = f(E_{\scriptscriptstyle UCN}) \left( \frac{vA}{4V} \right).
\end{equation}

Generally, the hydrogen in the organic materials has a large up-scattering cross section for the UCNs. Therefore the hydrogen in contact with UCN must be replaced with deuterium. The inner wall of the storage volume needs to be coated with a layer of deuterated polystyrene (dPS), whose Fermi potential is about 160~neV.\cite{Bodek08} The TPB converter coating also needs to be deuterated. These technologies are under development and tests by the SNS nEDM collaboration.\cite{Leung20} With a loss probability per bounce of $f(E_{\scriptscriptstyle UCN}) = 10^{-5}$, the same requirement as the SNS nEDM UCN storage cells \cite{SNS_nEDM19}, the time constant due to wall loss $\tau_{wall}$ is about 1672.0~s. \iffalse {\color{blue} (\missingcommand{tau_wall})}\fi

$\tau_{up}$ is the loss rate due to upscattering of neutrons by quasi-particles, phonons and rotons, in superfluid helium. It is greatly suppressed by a Boltzmann factor. At $T<0.6$~K, the dominant process is multi-phonon scattering following
\begin{equation} \label{eqn:tau_up_UCN}
    \frac{1}{\tau_{up}} = \frac{T^7}{100}~[\mbox{s}^{-1}].
\end{equation}
At $T=0.5$~K, $\tau_{up} = 12800$~s. With all the contributions above included in Eqn. (\ref{eqn:tau_tot_UCN}), the storage time constant $\tau_{tot}$ is about 144.2~s. \iffalse {\color{blue} (\missingcommand{tau_tot})}\fi This is merely an estimate. $\tau_{tot}$ will be accurately measured in experiments in order to acquire a high-precision determination on the $^3$He concentration $X_{\scriptscriptstyle He3}$ in Subsection \ref{subsec:helium3_meas}. 

\subsection{UCN Production}
The UCNs for this measurement are produced {\it in situ} in the neutron decay volume by super-thermal process: a 8.9~\AA~beam of CNs are down-scattered inelastically into UCNs via exciting a single phonon in superfluid helium.\cite{Golub77} The UCN density can build up in the decay volume with the time constant $\tau_{tot}$. The accumulated UCN density in liquid helium exposed to the CN beam is given by

\begin{equation}
    \rho_{\scriptscriptstyle UCN}(t_{\scriptscriptstyle fill})=R\tau_{tot} \left[ 1-\exp{\left( -\frac{t_{\scriptscriptstyle fill}}{\tau_{tot}}\right)} \right].
\end{equation}

$\tau_{tot}$ is the storage time constant as given in Eqn. (\ref{eqn:tau_tot_UCN}), and the production rate per unit volume $R$ is given by 

\begin{equation}
    R=2.2\times10^{-8}\left(  \frac{d\Phi}{dE}\right)  [\mbox{cm}^{-3}\mbox{s}^{-1}]
\end{equation}

where an incident flux spectrum of $\left(d\Phi/dE\right)$ is in units of (cm$^{-2}$~s$^{-1}$~\AA $^{-1}$), and the production of UCNs is up to the maximum storage Fermi energy of 160~neV.\cite{Golub77, BAKER03} As shown in Fig. \ref{fig:fig_FNPB}, the 8.9~\AA~CN flux in the SNS FnPB is about $5.9\times10^7 $~Hz~cm$^{-2}$~\AA$^{-1}$. \iffalse {\color{blue} (\missingcommand{UCN_dPhi_dt_dlambda})}\fi The production rate per unit volume is deduced to be $R \approx 1.3$~UCNs cm$^{-3}$~s$^{-1}$. So the steady state UCN density $\rho_{\scriptscriptstyle UCN}$ can reach an average of about 141.4 UCNs cm$^{-3}$ \iffalse {\color{blue} (\missingcommand{rho_UCN_fill})}\fi with 200~s of beam filling. {\it i.e.} a total number $N_0\approx7.5\times10^4$ \iffalse {\color{blue} (\missingcommand{N_UCN_fill})}\fi of UCNs can be filled in the storage cell with a beam-occupied volume of 530.1 cm$^3$. \iffalse {\color{blue} (\missingcommand{V_beam})}\fi However, the initial UCN filling number $N_0$ varies in different runs depending on the CN beam intensity and stability, and is of little use in the data analysis due to the large uncertainty. The estimate above is for the sake of presenting the order of magnitude.

\subsection{A Shortcut to Examine Neutron Lifetime Enigma}
The observed capture and decay rates of UCNs in the storage volume as well as their ratio are given by

\begin{equation}
    \dot{N}^{\scriptscriptstyle (p+t)}_{\scriptscriptstyle UCN}(t) 
    = -\epsilon'_{\scriptscriptstyle He3} \frac{N_0}{\tau_{\scriptscriptstyle He3}} \exp{\left(- \frac{t}{\tau_{tot}} \right)}, \label{eqn:UCN_captRate} 
\end{equation}

\begin{equation}
    \dot{N}^{\scriptscriptstyle (\beta)}_{\scriptscriptstyle UCN}(t) 
    = -\epsilon'_{\beta} \frac{N_0}{\tau_{\beta}} \exp{\left(- \frac{t}{\tau_{tot}} \right)},  \label{eqn:UCN_betaRate}
\end{equation}

\begin{equation} \label{eqn:ratio_capture_beta_UCN}
    \Tilde{\kappa}_{\scriptscriptstyle UCN} 
    = \frac{\dot{N}^{\scriptscriptstyle (p+t)}_{\scriptscriptstyle UCN}(t_1)}{\dot{N}^{\scriptscriptstyle (\beta)}_{\scriptscriptstyle UCN}(t_2)} 
    = \frac{\epsilon'_{\scriptscriptstyle He3}}{\epsilon'_{\beta}} \frac{\tau_{\beta}}{\tau_{\scriptscriptstyle He3}} \exp{\left(- \frac{t_1-t_2}{\tau_{tot}} \right)},
\end{equation}

where $\epsilon'_{\scriptscriptstyle He3}$ and $\epsilon'_{\beta}$ are the detection efficiencies of UCN capture and decay events that can be obtained via simulation. Since the $^3$He capture events can be distinguished from the decay events via the difference in the $1/t$ decay rate of after-pulses, $\dot{N}^{\scriptscriptstyle (p+t)}_{\scriptscriptstyle UCN}(t)$ and $\dot{N}^{\scriptscriptstyle (\beta)}_{\scriptscriptstyle UCN}(t)$ can be directly acquired in the measurement. The ratio $\Tilde{\kappa}_{\scriptscriptstyle UCN}$ can then be calculated as in Eqn. (\ref{eqn:ratio_capture_beta_UCN}), by counting both event rates in the identical time bins, {\it i.e.} $t_i=t_1=t_2$, 

\begin{equation}
\label{eqn:kappa_UCN}
    \Tilde{\kappa}_{\scriptscriptstyle UCN}(t_i) 
    = \frac{\epsilon'_{\scriptscriptstyle He3}}{\epsilon'_{\beta}} \frac{\tau_{\beta}}{\tau_{\scriptscriptstyle He3}} 
    = \frac{\epsilon'_{\scriptscriptstyle He3}}{\epsilon'_{\beta}} \tau_{\beta} n_{\scriptscriptstyle He3} \sigma_{\scriptscriptstyle He3}^{th}v_{\scriptscriptstyle He3}^{th}.
\end{equation}

$\Tilde{\kappa}_{\scriptscriptstyle UCN}(t_i)$ should statistically fluctuate around its true value. In this scenario, the neutron lifetime enigma can be examined by simply comparing $\tau_{\beta} n_{\scriptscriptstyle He3}$ obtained above by Eqn. (\ref{eqn:kappa_UCN}) in the UCN storage volume with that by Eqn. (\ref{eqn:nTauEqu}) in the beam decay volume. The real value of $^3$He density $n_{\scriptscriptstyle He3}$ is no longer necessary. But there is a caveat: the success of this "shortcut" trick greatly depends on how well one can characterize the background events, especially the Compton events induced by the gamma rays, and separate them from decay events through analysis and modelling. 

\subsection{Characterization on the Purity of Helium-4}

The production of isotopically pure $^4$He can be carried out in a purifier similar to that designed by Hendry and McClintock.\cite{McClintock87} For a residual $^3$He concentration, $X^{\scriptscriptstyle (0)}_{\scriptscriptstyle He3}<0.1\%X_{\scriptscriptstyle He3}\approx2\times10^{-13}$, the $^3$He capture time constant $\tau^{\scriptscriptstyle (0)}_{\scriptscriptstyle He3}$ is expected to be more than $2\times10^5$~s. \iffalse {\color{blue} (\missingcommand{tau_0_3})}\fi With the property of the UCN storage volume described in Subsection \ref{subsec:ucn_storage}, the total storage time $\tau^{\scriptscriptstyle (0)}_{tot}$ is expected to be about 551.7~s, \iffalse {\color{blue} (\missingcommand{tau_0_tot})}\fi and the UCN density $\rho^{\scriptscriptstyle (0)}_{\scriptscriptstyle UCN}$ can reach about 219.2 UCNs cm$^{-3}$. \iffalse {\color{blue} (\missingcommand{rho_0_UCN_fill})}\fi With 200~s of beam filling, it may achieve a fill of $N^{\scriptscriptstyle (0)}_0\approx1.16\times10^5$ \iffalse {\color{blue} (\missingcommand{N_0_UCN_fill})}\fi UCNs in the storage volume.

Because of the scarcity of the $^3$He atoms in the isotopically pure $^4$He liquid, a total number of neutron capture events $\Delta N^{\scriptscriptstyle (p+t,0)}_{\scriptscriptstyle UCN}$ may be counted over a period of storage time $\Delta t$ via identification on the after-pulses of scintillation events. So is the total number of the neutron decay events $\Delta N^{\scriptscriptstyle (\beta,0)}_{\scriptscriptstyle UCN}$. Integration on Eqns. (\ref{eqn:UCN_captRate}) and (\ref{eqn:UCN_betaRate}), respectively, gives

\begin{equation}
    \Delta N^{\scriptscriptstyle (p+t,0)}_{\scriptscriptstyle UCN} 
    =- \epsilon'_{\scriptscriptstyle He3} N^{\scriptscriptstyle (0)}_0\frac{\tau^{\scriptscriptstyle (0)}_{tot}}{\tau^{\scriptscriptstyle (0)}_{\scriptscriptstyle He3}} \left[ 1-\exp{\left(- \frac{\Delta t}{\tau^{\scriptscriptstyle (0)}_{tot}} \right)} \right],
\end{equation}

\begin{equation}
    \Delta N^{\scriptscriptstyle (\beta,0)}_{\scriptscriptstyle UCN} 
    =- \epsilon'_{\beta} N^{\scriptscriptstyle (0)}_0\frac{\tau^{\scriptscriptstyle (0)}_{tot}}{\tau_{\beta}} \left[ 1-\exp{\left(- \frac{\Delta t}{\tau^{\scriptscriptstyle (0)}_{tot}} \right)} \right].
\end{equation}

With the ratio of two equations above, the residual $^3$He concentration, $X^{\scriptscriptstyle (0)}_{\scriptscriptstyle He3}$, is obtained via the following equation,

\begin{equation}
\label{eqn:he3_residual}
    2.56\times10^{7}X^{\scriptscriptstyle (0)}_{\scriptscriptstyle He3}~[\mbox{s}^{-1}] 
    = \tau^{{\scriptscriptstyle (0)}~ -1}_{\scriptscriptstyle He3}
    =  \frac{1}{\tau_{\beta}} \frac{\epsilon'_{\beta}}{\epsilon'_{\scriptscriptstyle He3}} \frac{\Delta N^{\scriptscriptstyle (p+t,0)}_{\scriptscriptstyle UCN}}{\Delta N^{\scriptscriptstyle (\beta,0)}_{\scriptscriptstyle UCN}}.
\end{equation}

It is expected the counts of background Compton events might be comparable to or even larger than that of capture events with the residual $^3$He concentration. It undoubtedly introduces a large uncertainty on $X^{\scriptscriptstyle (0)}_{\scriptscriptstyle He3}$, but may not appear as large after propagated to that of the measured $X_{\scriptscriptstyle He3}$ in the later prepared helium mixture. The same applies to the value of $\tau_{\beta}$ used in Eqn. (\ref{eqn:he3_residual}). Despite simply fed with the PDG value, the error should be negligible after propagated to that of $X_{\scriptscriptstyle He3}$. 

Meanwhile, since $\tau^{{\scriptscriptstyle (0)}~ -1}_{\scriptscriptstyle He3}$ is negligible, the storage time $\tau^{\scriptscriptstyle (0)}_{tot}$ in isotopically pure $^4$He is expressed as

\begin{equation} \label{eqn:tau_tot_UCN_pureHe4}
    \tau^{{\scriptscriptstyle (0)}-1}_{tot} = \tau_{\beta}^{-1}+\tau_{up}^{-1}+\tau_{loss}^{-1}.
\end{equation}

$\tau^{\scriptscriptstyle (0)}_{tot}$ can be accurately measured by fitting the time-dependent decline of $\dot{N}^{\scriptscriptstyle (\beta)}_{\scriptscriptstyle UCN}(t)$ as in Eqn. (\ref{eqn:UCN_betaRate}) with normalization to an arbitrary time zero after the beam stops. The uncertainty of $\tau^{\scriptscriptstyle (0)}_{tot}$ greatly depends on a good understanding of the Compton event rate. 

\subsection{Accurate Determination of Helium-3 Concentration in the Helium Mixture}
\label{subsec:helium3_meas}
Once the isotopically pure $^4$He is characterized, the desired $^3$He concentration of about $X_{\scriptscriptstyle He3}=2\times10^{-10}$ can be prepared by mixing with natural helium of a known $^3$He abundance as to a preset volume ratio. The $^3$He atoms may be expelled from the volume via heat flush to fine tune its concentration.\cite{McClintock87,Baym15} 
It is noteworthy that Eqns. (\ref{eqn:ratio_capture_beta_UCN}) and (\ref{eqn:he3_residual}) are equivalent but neither can be used in the accurate determination of $X_{\scriptscriptstyle He3}$, because they proportionally link $\tau_{\beta}$ to $\tau_{\scriptscriptstyle He3}$. Therefore, $X_{\scriptscriptstyle He3}$ should be extracted from the difference between $\tau^{\scriptscriptstyle (0)}_{tot}$ in the isotopically pure $^4$He and $\tau_{tot}$ in the prepared helium mixture, according to the following expression,

\begin{equation}
\label{eqn:he3_accurate}
    2.56\times 10^{7}X_{\scriptscriptstyle He3}~[\mbox{s}^{-1}] 
    {\color{black}= n_{\scriptscriptstyle He3} \sigma_{\scriptscriptstyle He3}^{th} v_{\scriptscriptstyle He3}^{th} }
    = \tau_{\scriptscriptstyle He3}^{-1} = \tau_{tot}^{-1} - \tau^{{\scriptscriptstyle (0)} -1}_{tot} .
\end{equation}

In a storage measurement, both the decay and capture event rates, $\dot{N}^{\scriptscriptstyle (p+t)}_{ \scriptscriptstyle UCN}(t)$ and $\dot{N}^{\scriptscriptstyle (\beta)}_{ \scriptscriptstyle UCN}(t)$, decline with an identical time dependence on the total storage time constant $\tau_{tot}$ of the volume, as in Eqns. (\ref{eqn:UCN_captRate}) and (\ref{eqn:UCN_betaRate}). $\tau_{tot}$ can be measured via both channels independently, though the decay rate channel might have a higher uncertainty due to the contamination of background Compton events. The detection efficiencies, $\epsilon'_{\scriptscriptstyle He3}$ and $\epsilon'_{\beta}$, are irrelevant 
{\color{black} as long as they remain stable in one measurement cycle. However, it has been experimentally found the UCN loss rate in a material bottle storage is not completely described by an exponential decline as in Eqns. (\ref{overview10}), (\ref{eqn:UCN_captRate}) or (\ref{eqn:UCN_betaRate}). A dual exponential fit often works much better than the single exponential fit in a storage volume.\cite{Bodek08} It is possibly due to the change of UCN spectrum during the storage, especially as UCNs of higher energy have more chances to be up-scattered and absorbed on the wall, or to leak out at some spots of wall materials with lower Fermi potentials. As a result, it may impose systematic uncertainties on the measurement of both $\tau_{tot}$ and $\tau^{\scriptscriptstyle (0)}_{tot}$, which is tightly related to the energy spectrum of the UCNs generated in the storage volume and quality of the wall material.}

{\color{black} Nevertheless, $X_{\scriptscriptstyle He3}$ given by Eqn. (\ref{eqn:he3_accurate}) is not a good option to calculate the beam neutron lifetime via Eqn. (\ref{eqn:nTauEqu}), because the prefactor $2.56\times 10^{7}$ has inherited an uncertainty of 0.13\% from $\sigma_{\scriptscriptstyle He3}^{th}$. In fact, Eqn. (\ref{eqn:nTauEqu}) asks for the combined term $n_{\scriptscriptstyle He3} \sigma_{\scriptscriptstyle He3}^{th} v_{\scriptscriptstyle He3}^{th}$, which can be directly acquired once and for all by Eqn. (\ref{eqn:he3_accurate}). Therefore, $\sigma_{\scriptscriptstyle He3}^{th}$ by itself is no longer in the path of final calculation of the beam neutron lifetime, and hence, its uncertainty becomes irrelevant. In the end, the sub-0.1\% accuracy of the beam neutron lifetime relies on adequate statistics acquired for both $\tau_{tot}$ and $\tau^{\scriptscriptstyle (0)}_{tot}$, as well as the ratio of event rates $\kappa =\dot{N}_{p+t}/\dot{N}_{\beta}$ discussed earlier in Subsection \ref{subsec:ratio_capt_decay}.}

\subsection{Gamma Ray Backgrounds in the UCN Storage Measurement}
\label{subsec:UCN_gamma}

Similarly, gamma rays generated from the neutron captures on the surrounding materials are the major contributor to the background, but the UCN storage measurement is mostly sensitive to the delayed components rather than the prompt. In addition to the capture and scattering by the windows, the 8.9~\AA~CN beam has a scattering cross section of about 0.025~barns on liquid helium at 0.5~K.\cite{Sommers55,Abe01}  4\% of CNs will be inelastically scattered by the phonons in liquid helium. A very small portion is down-converted into UCNs and trapped in the storage volume, whereas majority of the scattered neutrons project at angles of around 84 degrees off the incident direction. About 12\% of the scattered neutrons are captured by hydrogen, on top of those captured by the diamond window. The resultant prompt gamma rays are intensive but vanish right after the beam halts. Therefore, it doesn't affect the measurement of UCN storage time $\tau_{tot}$. More troublesome are the delayed gamma rays from the 0.002\% of scattered neutrons captured on the fluorine in the 50~$\mu$m thick PTFE reflector. After 200~s of UCN filling with the 8.9~\AA~CN beam, the neutron-activated fluorine saturates. The resultant delayed gamma rays of 1.6~MeV induce Compton scintillation at 2.5~Hz \iffalse {\color{blue} (scaled)}\fi in liquid helium and 1.7~Hz \iffalse {\color{blue} (scaled)}\fi in the fibers at time-zero when the beam turns off. Since the half-life of the activated fluorine is 11.16~s, the rate of Compton scintillation in liquid helium drops below 1~Hz after 15~s. \iffalse {\color{blue} (scaled)}\fi At the meantime, the rate of decay events is 76.8~Hz, \iffalse {\color{blue} (scaled)}\fi assuming the storage time is 144.2~s and the initial UCN number is $N_0\approx7.5\times10^4$. The delayed gamma ray background discussed above can be eliminated by replacing the PTFE reflector with a polymer reflector, such as Vikuiti VM2000 \cite{reflector}, but others may still remain due to impurities in the surrounding materials. 

% 8.9A CN beam of 4.2e+08 Hz scattered rate by LHe: 1.68e+07 Hz
% CN capture rate: cell+fiber=0.1186, holder=0.0020 for 5mm thick PTFE
% CN capture rate: cell+fiber=0.1187, holder=0.0002 for 0.5mm thick PTFE
% CN cpature rate: cell+fiber=0.1187, holder=0.00002 for 0.05mm or 50um thick PTFE
% Delayed Gamma Rays of 1.6MeV turn into Compton scintillation in LHe: 503.47 Hz (5mm PTFE) --> 5.0Hz (50um PTFE)
% Delayed Gamma Rays of 1.6MeV turn into Compton scintillation in fiber: 330.17 Hz (5mm PTFE) --> 3.3Hz (50um PTFE)
% scale the values by 1/2 because the 8.9A CN beam is modified with a flux 2.1e8 Hz.
% half life 11.16s == log lifetime 16.1s

\section{Conclusion}
This proposed experiment has great potential to reach a sensitivity of 0.1\% or sub-1 second in neutron lifetime measurement. It offers an entirely different set of systematic uncertainties from the existing beam experiments. 
{\color{black} As explicitly expressed in Eqn. (\ref{eqn:nTauEqu}), the calculation of $\tau_{\beta}$ is independent of the neutron flux and the geometry of the decay volume.}
Most uniquely, it does not require any magnetic field, and may be set up to test the hypothesis of neutron-mirror neutron $n-n'$ oscillations, where the intensity of magnetic field plays an important role.\cite{Mirror19_1, Mirror19_2, Mirror06_1, Mirror06_2} Its apparent disadvantage is the flux of a CN beam at the wavelength $\lambda > 16.5$~\AA~being much weaker than that of the most commonly in use 4--5~\AA~CN beams. It will take more beam time to gain adequate statistics. There might be a possibility to optimize the CN beam output for the long wavelength specifically needed in this experiment. Nevertheless, the low event rate allows a thorough characterization of the temporally spaced events with less interference. 
{\color{black} High precision measurement is required on the three quantities: the ratio of event rates $\kappa$ in the CN decay volume, as well as the storage time constants $\tau_{tot}$ with liquid helium mixed with a proper concentration of helium-3, and $\tau^{\scriptscriptstyle (0)}_{tot}$ with isotopically pure liquid helium-4 in the UCN storage volume. Both the volumes can be connected and share the same batch of prepared liquid helium, where statistics on two measurements can be gained simultaneously. The 8.9~\AA~CN beam can be extracted from the main beam by a monochromator \cite{Fomin15, Mattoni04} and sent to the UCN storage volume, whereas the 17~\AA~CNs can be selected by a set of double or triple choppers in the main beam and sent to the CN decay volume. Hopefully, a good design of the double or triple choppers can effectively filter the contamination of neutrons in the unwanted spectrum. The Atom Trap Trace Analysis can be set up to characterize the helium-3 concentration in samples as a verification of the result offline.} Furthermore, while the intrinsic UCN storage time $\tau^{\scriptscriptstyle (0)}_{tot}$ is being measured in the storage volume filled with isotopically pure liquid $^4$He, the neutron beta decay spectrum can be simultaneously obtained in the decay volume at a good resolution and accuracy with this detector. It may provide a measurement on the Fierz interference term $b$ in the energy dependent neutron beta decay rate at a potentially high precision compared to \iffalse{\color{blue} $b_n = 0.025 \pm 0.019$, the weighted average of}\fi the most recent results \cite{Fierz19_1, Fierz19_2}.  
{\color{black} This paper only covers the proof of principle for the proposed experiment. Many unknown technical issues will unsurprisingly emerge as the engineering designs and prototyping tests proceed. We hope it can eventually help resolve the neutron lifetime enigma.}

% Fierz interference term ==> Sun19: $0.066 \pm 0.0475 = 0.066 \pm 0.041_{stat} \pm 0.024_{syst}$ vs Saul19: $b = 0.017(21)$

\section*{Acknowledgement}
This paper greatly benefits from the publications, notes and data shared by the SNS nEDM collaboration. The author acknowledges Vince Cianciolo, Bradley W. Filippone, Roy J. Holt, Humphrey J. Maris, Jeffrey S. Nico, George M. Seidel, Christopher M. Swank, Fred E. Wietfeldt and Liyuan Zhang for help and discussions on many topics presented in this paper. This work is supported in part by the National Science Foundation under Grant No. 1812340.

\end{document}